\documentclass[aps,eqsecnum,amsmath,amssymb,twocolumn]{revtex4}
\usepackage{graphics, setspace}
\usepackage[letterpaper, dvips,width=7.5in,height=8.5in,includemp=false]{geometry}
\usepackage[vcentermath]{youngtab}

\newcommand{\be}{\begin{equation}}
\newcommand{\ee}{\end{equation}}

\begin{document}

\title[title]{The collective quantization of three-flavored Skyrmions revisited}
\author{Aleksey Cherman}
\email{alekseyc@physics.umd.edu}

\author{Thomas D. Cohen}
\email{cohen@physics.umd.edu}

\author{Timothy R. Dulaney}
\email{tdulaney@umd.edu}

\author{Erin M. Lynch}
\email{elynch@umd.edu}

\affiliation{Department of Physics, University of Maryland,
College Park, MD 20742-4111}

\begin{abstract}
A self-consistent large $N_c$ approach is developed for the
collective quantization of $SU(3)$ flavor hedgehog solitons, such as
the Skyrmion.  The key to this analysis is the determination of all
of the zero modes associated with small fluctuations around the
hedgehog. These are used in the conventional way to construct
collective coordinates. This approach differs from previous work in
that it does not implicitly assume that each static zero mode is
associated with a dynamical zero mode. It is demonstrated explicitly
in the context of the Skyrmion that there are fewer dynamical zero
modes than static ones due to the Witten-Wess-Zumino term in the
action. Group-theoretic methods are employed to identify the
physical states resulting from canonical quantization of the
collectively rotating soliton.  The collective states fall into
representations of $SU(3)$ flavor labeled by $(p,q)$ and are given
by $(2J, \frac{Nc}{2} -J)$ where $J=\frac{1}{2},\frac{3}{2},\cdots $
is the spin of the collective state. States with strangeness $S > 0$
do not arise as collective states from this procedure; thus the
$\theta^{+}$ (pentaquark) resonance does not arise as a collective
excitation in models of this type.
\end{abstract}

\maketitle

\section{Introduction}

The collective quantization of $SU(3)$ flavor hedgehog soliton
models ({\it eg.} the Skyrme model) was first implemented more than two
decades ago \cite{Guadagnini,hedgehogs}.  The approach appeared to
be a natural extension of the collective quantization of two
flavored Skyrmions introduced by Adkins, Nappi and Witten (ANW)
\cite{Adkins:1983ya} with important constraints imposed due to the
Witten-Wess-Zumino term.  The results of this approach had numerous
phenomenological successes in the description of the octet and
decuplet baryons.  However, as was noticed long ago by Praszalowicz
\cite{Praszalowicz:2005ps} and emphasized more recently by Diakonov,
Petrov and Polyakov (DPP) \cite{Diakonov:1997mm} such a procedure
also predicts the existence of collective states with exotic quantum
numbers such as a strangeness $S=+1$ baryon.  Since chiral soliton
models are believed to correctly encode the large $N_c$ behavior of
QCD, the prediction of exotic baryon states in these models appears
to imply that QCD, at least at large $N_c$, predicts the existence
of exotic baryons.

This prediction appeared to assume great significance with the
initial reports of the detection of an exotic $S=+1$ baryon, the
$\theta^+$, with a mass of $\sim 1540$ MeV \cite{pentafirst} and a small
width.  Subsequently, multiple claims confirming this state
\cite{pentaexp} added weight to the prediction. The mass claimed in these
experimental papers was in remarkably good agreement with the mass
predicted using the method of Refs. \cite{Diakonov:1997mm}; the widths were consistent.

However, the situation is far more complex.  On the experimental
side the evidence for the $\theta^+$ is based on relatively
low-statistics experiments with significant cuts on the data.
Numerous experimental groups, including
many with much higher statistics, did not see the $\theta^+$.  Particularly significant are
the two recent high statistics experiments reported by CLAS, which
found no evidence for the $\theta^+$ and were done at similar
dynamics to previous experiments which claimed to observe the
resonance \cite{clas1}. As a result of this there is considerable
skepticism in the community as to whether the $\theta^+$ is real.

The theoretical situation is also complicated.  Although the
predictions of the octet and decuplet baryons from the collective
quantization of the $SU(3)$ soliton appears to be sound, it has been argued
that the predictions of exotic states such as the $\theta^+$
are not consistent with the underlying large $N_c$
dynamics \cite{Cohen:2003yi, Cohen:2003mc, Itzhaki:2003nr, Pobylitsa:2003ju}.

The crux of the issue is the use of the rigid rotator scheme to
quantize the hedgehog solitons.  The rigid rotator approach was
developed by ANW for the $SU(2)$ Skyrme model. It is generally
agreed that this method correctly describes the large $N_c$
collective dynamics of the Skyrme model, although it must be
modified for models with explicit quarks or vector meson degrees of
freedom\cite{Cohen:1986va}. The quantization of $SU(3)$ solitons in
Refs. \cite{Guadagnini,hedgehogs} borrows the ANW method directly
and the predictions of exotic states by Praszalowicz
\cite{Praszalowicz:2003ik} and DPP \cite{Diakonov:1997mm} depend on
this assumption.

However, the rigid rotator approach for exotic states of $SU(3)$
hedgehog models is known to be inconsistent with the large $N_c$
dynamics used to justify the
models\cite{Cherman:2004qx,Cohen:2003yi,Cohen:2003mc,Pobylitsa:2003ju}.
This can be seen in a number of indirect ways.  Perhaps most
compelling is the fact that collective excitations have a level
spacing which goes to zero at large $N_c$, corresponding to slow
classical motion which decouples from the intrinsic degrees of
freedom. However, the excitation energy of the exotic states is of order
$N_c^0$.  Initially this was controversial; Diakonov and Petrov made
several counterarguments justifying the rigid rotor approach
\cite{Diakonov:2003ei}.  However these were shown to be
invalid\cite{Cohen:2003mc} and in a recent paper Diakonov and Petrov
now agree that the approach is invalid at large $N_c$
\cite{Diakonov:2005ib}.

While the controversy regarding whether exotic states appear as
collective states of the soliton model appears to be resolved, an
important theoretical question remains open: how \emph{does} one
consistently quantize these models?  There is an obvious {\it ad
hoc} approach --- simply use the procedures and results of the
rigid rotor quantization of Refs. \cite{hedgehogs,Guadagnini}, but
restrict their application to the non-exotic states of the theory.
Such an approach has one cardinal virtue --- {\it a priori} its
results are virtually certain to be correct at large $N_c$.  The
reason for this is simply that in implementing this scheme one
produces exactly the same collective spectrum as emerges from a
model-independent analysis based on large $N_c$ consistency rules
\cite{dashen} which, not coincidentally, turns out be identical to
that of a large $N_c$ constituent quark model.

However, this {\it ad hoc} approach is deficient in two ways.
First, from a purely theoretical perspective, making such an {\it
ad hoc} prescription is unsatisfactory.  One needs to understand
what is wrong with the rigid rotor analysis and how to do it
correctly. Second, from a more pragmatic perspective, if one
wishes to treat these models in a consistent $1/N_c$ expansion
beyond leading order, it is essential to do the bookkeeping
correctly by separating the collective degrees of freedom from the
non-collective ones in a transparent manner.

As will be discussed below, the key issue turns out to be the one
identified in Refs. \cite{Cohen:2003mc,Cherman:2004qx}; namely,
the correct identification of zero modes in the spectrum of small
fluctuations around the soliton.  As is standard in soliton
physics, the degrees of freedom associated with zero modes get
promoted to collective degrees of freedom which decouple from the
other degrees of freedom in the problem in the semiclassical limit
(large $N_c$ in the present context) . There is an important
subtlety for $SU(3)$ solitons which is not present for $SU(2)$
solitons; namely, that there are fewer dynamic zero modes than
static ones.  The distinction between dynamic and static zero
modes will be discussed in the following section.

Before embarking on this analysis we should note that at a formal
level this approach is based on exact $SU(3)$ flavor symmetry.
Any attempt to connect this to the real world, where there is a
nontrivial level of $SU(3)$ breaking, requires an additional
perturbative treatment in the $SU(3)$ breaking term.  There is an
alternative formalism due to Callan and Klebanov
\cite{Callan:1985hy} which encodes the $SU(3)$ violations from the
outset, and in which fluctuations which change the strangeness are
always regarded as non-collective vibrations. We make no claims
here regarding the relative efficiency of the two approaches.
Indeed we fully expect the approaches to be essentially equivalent
for small quark masses. Nevertheless it is important at least at
the level of mathematical physics to understand how to
collectively quantize $SU(3)$ solitons.

While the general approach discussed here can be used for any $SU(3)$
chiral soliton model at large $N_c$, we focus on a simple variant of
the Skyrme model in order to keep the discussion focused.

The plan of this paper is as follows:  In the next section we
review a few basic issues about zero modes and collective
quantization in soliton physics. There we  focus on the
circumstances in which rigid rotations are valid, and discuss how
to collectively quantize the theory when rigid rotation fails.  We
also make the distinction between static and dynamic zero modes
and show how there can be more of the former than of the latter.
Following this is a section introducing the  Skyrme model.  Next,
we focus on identifying the flavor zero-modes of the hedgehog
Skyrmion solution, and show that the system has only ten zero
modes (seven static and three dynamic ones) rather than fourteen
(seven static and seven dynamic modes), as is implicitly assumed
in rigid-rotator quantization. We derive a collective Hamiltonian,
and describe the states that arise in the quantization of the
resulting collective coordinates.  These turn out to be precisely
those obtained in the model-independent large $N_c$ approach of Ref.
\cite{dashen}.

\section{Zero modes and the collective quantization of solitons }

\subsection{Collective quantization: Generalities}
There is a standard procedure for quantizing solitons
\cite{Jackiw:1977yn,Rajaraman}. Suppose we have a quantum
field-theoretic Lagrangian that, when treated classically, admits
solitonic solutions.   For simplicity, in this discussion we will
restrict our attention to theories such as Skyrme's original
model\cite{Skyrme:1961vq} in which all fields are Lorentz
scalars. Theories which include vector or spinor degrees of
freedom add some minor technical complications which make the
argument a bit less transparent, but do not alter the essential
details of the problem. The solitonic solutions can be interpreted
as the first approximations to the quantum ground state of the
system in a semiclassical expansion. In the present context this
expansion turns out to be equivalent to the $1/N_c$ expansion.
Small oscillations around the classical soliton are then
interpreted as corrections to this approximation. Zero modes in
the fluctuation spectrum turn out to play a central role in the
analysis.

The Lagrangian necessarily has continuous symmetries which are
broken by the classical soliton configuration.  While the
Lagrangian is Poincar\'e invariant, the soliton breaks
translational invariance. Other symmetries of the Lagrangian may
also be broken. By applying symmetry operations at the classical
level it is therefore possible to continuously change a solitonic
configuration to obtain distinct solitonic configurations of the
same energy. A small fluctuation in an energetically flat
direction is a zero mode --- there is no restoring force.  It
should be  apparent that these classical soliton configurations
cannot directly correspond an energy eigenstate of the quantum
system; the eigenstates have well defined values of the symmetry
operators while the classical solitons do not.  Thus, for example,
the fact the soliton breaks translational invariance means that
the classical soliton cannot correspond to a quantum state with
good momentum.  To make a connection with the physical states, it
thus becomes essential to quantize the degrees of freedom
associated with the symmetry breaking, which, as we have seen, are
the zero modes.  These are denoted the collective degrees of
freedom.  If one is working in the extreme semiclassical regime
({\it eg.} at large $N_c$) the collective degrees of freedom
should completely decouple from the remaining degrees of freedom.
In this limit it is possible to quantize them separately by
imposing the appropriate commutation relations.

\subsection{Static and dynamic zero modes}
\label{gen}

Formally one can find the zero modes by solving the eigenvalue
equation for small classical harmonic fluctuations of the fields
around the static soliton.  The zero modes are so-named because they
have zero frequency.  Generally the equation for small amplitude
fluctuations is second order in time.  The most general fluctuation
equation is of the form
 \begin{equation}
\delta \ddot{ \phi}= \mathbb{A}\delta \phi + \mathbb{B} \delta
\dot{ \phi},
 \label{geneq}\end{equation}
where the $\delta \phi $ are the first order fluctuations of the
fields and $\mathbb{A}, \mathbb{B}$ are differential operators in
space. If there are no terms in the action linear in the time
derivatives of the fields, then $\mathbb{B}=0$ in Eq.~(\ref{geneq}).
Harmonic fluctuations with finite frequency are given by a
generalized eigenvalue equation
 \begin{equation}
-\omega^2{\delta \phi}= \mathbb{A}\delta \phi - i \omega
\mathbb{B} {\delta \phi}.
 \label{geneig}\end{equation}

In addition to the finite frequency solutions of the preceding
form, there are zero modes. First consider the case $\mathbb{B}=0$
(no terms linear in the time derivative). In such a case the zero
modes necessarily come in pairs. Suppose there exists a static
field configuration $\psi$ which satisfies $\mathbb{A} \psi =0$.
Then $\delta \phi (t) = \psi$ and $\delta \phi (t)  = \psi t$ are
both solutions of the small fluctuation equation,
Eq.~(\ref{geneq}), but they do not correspond to finite frequency
harmonic oscillations; they are zero modes. Physically the first
of these two solutions corresponds to a static displacement of the
fields while the second corresponds to a ``kick'' in which the
fields receive an impulse. We will refer to these two type of
solutions as ``static'' and ``dynamic'' zero modes, respectively.

The distinction between static and dynamic zero modes is usually
not made.  For the situation described above in which there are no
terms in the action linear in the time derivatives and
$\mathbb{B}=0$ there is no need to distinguish them; they
necessarily come as a pair and it is sufficient to label the
entire pair as {\it the} zero mode.

Suppose, however, that there are terms in the action linear in the
time derivatives of the fields and, accordingly, $\mathbb{B} \neq
0$. Again, let us suppose that there exists a static field
configuration $\psi$ which satisfies $\mathbb{A} \psi =0$.  In
this case there will again be a static zero mode of the form
$\delta \phi (t) = \psi$ independent of the nature of
$\mathbb{B}$.  However, from the form of Eq.~(\ref{geneq}), it is
apparent that $\delta \phi (t)  = \psi t$ is not generically a
dynamic zero mode. Dynamical zero modes may be generated from the
static ones provided $\mathbb{B} \psi$ is either zero or
orthogonal to all zero modes of $\mathbb{A}$. In this case, it is
easy to see that the dynamical zero modes are of the form:
\begin{equation}
\label{dyn}
\delta \phi (t) =  \psi t + \Phi \; \; {\rm with} \; \; \mathbb{A}
\Phi = - \mathbb{B} \psi \, .
 \end{equation}
The shift $\Phi$ is well known in the context of many-body physics
where it is referred to as arising from``cranking''\cite{Cohen:1986va}.

Note, however, that Eq.~(\ref{dyn}) only has solutions when
$\mathbb{B} \psi$ is either zero or orthogonal to all the zero
modes of $\mathbb{A}$; if this is not true the condition
$\mathbb{A} \Phi = - \mathbb{B} \psi$ cannot be satisfied. Thus,
the number of static zero modes is the number of linearly
independent configurations satisfying $\mathbb{A} \psi =0$, while
the number of dynamic zero modes is the number of linearly
independent configurations simultaneously satisfying $\mathbb{A}
\psi =0$ and $\mathbb{B} \psi $ orthogonal to all the zero modes
of  $\mathbb{A}$. Thus, the number of dynamical zero modes is less
than or equal to the number of static modes. In considering the
collective quantization of three-flavored hedgehog models such as
the Skyrmion, we shall see that the number of dynamical zero modes
is less than the number of static zero modes; this fact will be
central to  the analysis.

To illustrate the difference between static and dynamic zero modes
more clearly, consider a simple example in classical mechanics.  Let
a particle with position $r$ move in $R^3$ with a potential $U$
(depending only on $z$) with a minimum at $z=0$. The Lagrangian is
\be
 L = \frac{1}{2} m (\partial_t \vec{r})^2 - U(\vec{r}), 
\;\;i = x,y,z
\;\;
\rm{and}
\;\; \partial_x U=\partial_y U=0 \;  
\ee 
and the equation of motion is simply 
\be 
\label{dumbEOM}
 m (\partial_{t} r_i)^{2} + \partial_i U(\vec{r})=0 . 
\ee
Clearly $\vec{r}=\vec{r}_0=(0,0,0)$ is a time-independent solution
to Eq. (\ref{dumbEOM}). The Lagrangian is invariant under
displacements in the $x-y$ plane. Thus, it is easy to see that
$\vec{r} = \vec{r}_0 + h\hat{x}$ is also a solution with the same
energy for any displacement $h$, and similarly for displacements in
the $\hat{y}$ direction.  Since any solution of Eq. (\ref{dumbEOM})
can be displaced time-independently in the $x,y$ directions without
any change in energy, we know that this system possesses two static
zero modes corresponding to displacements in the $(x,y,0)$ plane.

There are also solutions of the form $\vec{r}(t) = \vec{r}_0 + v t
\hat{x}$, where $v$ is a constant, as well as analogous solutions with
a term in the $\hat{y}$
direction.  Since $v$ can be arbitrarily small, and the energy is
quadratic in $v$, these solutions are essentially degenerate in
energy with the original one.  We identify the infinitesimal
fluctuation with finite velocity as the dynamic zero modes of the
system. With this identification, we see that in this system each
static zero mode is associated with a dynamic zero mode in the same
direction.

The reason for this association between static and dynamical zero
modes is the absence of any velocity-dependent forces in Eq.
(\ref{dumbEOM}). To see how such forces can break the association,
consider modifying the system above by giving the particle an
electric charge $e$, and adding a constant magnetic field
$\vec{B}=g \hat{z}$. It is apparent that the new system still
has two static zero modes corresponding to static displacements in
the $(x,y,0)$ plane.  However, there are no dynamic zero modes:
the magnetic field bends the particle into a closed orbit; the
frequency of this orbital motion is simply the cyclotron
frequency. Thus we see that a velocity-dependent term in the
equation of motion has removed the association between static and
dynamical zero modes, and that in presence of such a term, not
every static zero mode has a corresponding dynamical zero mode. In
terms of the framework given in Eqs. (\ref{geneq}) and
(\ref{dyn}), this situation is explained by the fact that
$\mathbb{B}$ is now non-zero due to the magnetic field, and the
condition $\mathbb{A} \Phi = - \mathbb{B}\psi$ cannot be
satisfied.

Let us now return to the soliton problem.  The motion associated
with the zero modes can occur classically at arbitrarily slow
speeds. At infinitesimally slow speeds these modes completely
decouple from all of the other degrees of freedom in the system.
If the expansion governing the semiclassical expansion (such as
the $1/N_c$ expansion) is valid, then this decoupling is valid
quantum mechanically up to corrections which are higher order in
the expansion. The quantum theory at lowest order is quite simple
conceptually: motion in the direction of the zero modes is
expressed in terms of a set of collective degrees of freedom;
appropriate commutation relations for these are imposed.  The
remaining degrees of freedom are harmonic to lowest order and are
quantized as harmonic degrees of freedom. Coupling between the
collective and non-collective (vibrational) degrees of freedom or
between distinct non-collective degrees of freedom are higher
order and can be treated perturbatively.

\subsection{Collective quantization}

Here we focus on the quantization of the collective degrees of
freedom.  One essential point is that these degrees of freedom are
associated with zero modes which in turn arise from the breaking of
symmetries at the classical level.  When these collective degrees of
freedom are quantized properly, the symmetries of the underlying
quantum system are thereby restored.

There are many equivalent ways to implement the quantization of
the collective degrees of freedom. For example, if explicit forms
for a classical collective coordinate and its conjugate momenta
can be found, then canonical quantization rules can be implemented
in a straightforward manner\cite{Jackiw:1977yn,Rajaraman}.
However, this may be cumbersome in many cases, particularly in
cases such as the Skyrmion where the soliton breaks multiple
symmetries with non-commuting generators.

There is a simple and general strategy for dealing with such
cases\cite{Cohen:1986va}. The procedure has three basic steps: i)
First, one obtains an explicit expression for the most general
classical collective motion ({\it i.e.}, a time-dependent field
configuration which solves the classical equation of motion for
arbitrarily slow motion and which is directed along a local zero
mode).  These classical configurations are uniquely specified by a
set of parameters; the number of parameters is equal to the number
of zero modes, both static and dynamic.  ii) Next, field-theoretic
expressions need to be obtained for the symmetry generators broken
by the classical soliton.  The insertion of the configuration
associated with collective motion of the soliton then gives an
explicit expression for the generators in terms of the parameters
which specify the collective motion.  iii)  The known commutation
rules for the quantum generators can only be satisfied if the
parameters specifying the collective motion are promoted to
quantum `collective' operators, the commutation relations of which
are derived from those of the known quantum operators.  By
imposing these derived commutation rules one quantizes the
collective motion in a manner consistent with the underlying
symmetries of the theory.

\section{The Skyrme Model}

In this section we briefly review the Skyrme model which we will
use to illustrate the issues of collective quantization.  As noted
in the Introduction, the methods are generally applicable, but it
is far more simple to discuss matters in the context of a specific
model.  Here we will use a Lagrangian identical in form to
Skyrme's original model \cite{Skyrme:1961vq} but generalized to
three flavors.   The Skyrme Lagrangian density is given by 
\be
\mathcal{L}_{S}=-\frac{f_{\pi}^2}{4} \mathrm{Tr}(L_{\mu} L^{\mu})
+ \frac{\epsilon^2}{4} \mathrm{Tr}([L_{\mu},L_{\nu}]^2).
\label{SK}
 \ee
 Here the left chiral current is defined as $L_{\mu} =
U^{\dagger}
\partial_\mu U$, with $U \in SU(3)$, and $U$ can be written as $U = e^{i \lambda_i
\phi^i/f_{\pi}}$, where the $\lambda^i$ are the Gell-Mann
matrices, and the $\phi^i$ are the Goldstone boson fields.

As noted by Witten \cite{Witten:1983tw}, a Lagrangian of the form
of Eq.~(\ref{SK}) does not correctly encode the anomaly structure
of QCD.  Moreover, no local term in a Lagrangian is capable of
doing this.  However, a nonlocal term in the action can. Thus, to
reproduce QCD's low energy anomalous structure, a term nonlocal in
$3+1$ dimensions (but local in $4+1$ dimensions) must be added to
the action.  This is the Witten-Wess-Zumino term:
 \be 
 \Gamma = \pm
\frac{i}{240 \pi^{2}} \int_{D_{5}^{\pm}} {d^{5}x \; \epsilon^{\mu
\nu \alpha \beta \gamma} \; \mathrm{Tr}(L_\mu L_\nu L_\alpha
L_\beta L_\gamma) }. 
\ee
 Here the boundary of $D_{5}^{\pm}= S^3\times S^1
\times [0,\pm 1]$ is compactified space-time.  (The sign ambiguity
is resolved by Stoke's theorem.)  The full action is then given by
\be
 \label{action} S = n \Gamma + \int{d^{4}x \; \mathcal{L}_S},
\;\; n\in \mathbb{Z} \;. 
\ee 
The restriction $n \in \mathbb{Z}$
was shown by Witten to emerge directly from a topological
consistency condition in an analysis formally analogous to Dirac's
electric charge quantization argument \cite{Witten:1983tx}.  To
reproduce QCD's anomaly structure one must take $n=N_c$. With this
addition the action above is known to encode the correct scaling
behaviors of large $N_c$ QCD, provided the parameters in Eq.
(\ref{action}) scale according to
 \be
 f_\pi \sim  N_c^{1/2}, \; \; \epsilon \sim N_c^{1/2} \, .
 \ee

The equations of motion of the model are obtained in the standard
way by varying the action.  A convenient way to express these is:
\be 
\label{eom}
 -\partial^{\mu} L_{\mu} - 2
\frac{\epsilon^2}{f_{\pi}^2} \partial^{\mu}
[L_{\nu},[L_{\mu},L^{\nu}]] + \frac{i N_c}{24 \pi^2 f_{\pi}^2}
\epsilon^{\alpha \beta \gamma \nu} L_{\alpha} L_{\beta} L_{\gamma}
L_{\nu}=0 . 
\ee Observe that the WWZ term is first order in the time
derivative in the equations of motion.

The equations of motion above admit topologically nontrivial
solitonic solutions.  To see what this means, note that if working
in $SU(2)$, the set of solution maps $U: S^3 \rightarrow S^3$
split into homotopically distinct classes, and we can associate a
winding number with these homotopy classes \cite{Zahed:1986qz}.
The winding number can be obtained from a current, 
\be
\label{baryonNum} B^\mu = i \frac{\epsilon^{\mu \nu \alpha
\beta}}{24 \pi^2}
 {\rm Tr}[ L_\nu L_\alpha L_\beta ],
 \ee
which is algebraically conserved: $\partial_\mu B^\mu = 0$.  The
winding number $B$ is simply the spatial integral of $B^0$. $B$ is
an integer valued function of the fields ($B: \pi_{3}(S^3)
\rightarrow Z$) and distinguishes between the homotopy classes.  The
generalization to SU(3) still yields a conserved current and integer
values of $B$. The current $B^\mu$ has a simple physical
interpretation---it represents the baryon current
\cite{Witten:1983tx}. Thus soliton solutions which have $B=1$
corresponds to baryons.

The form of the lowest energy solution to Eq. (\ref{SK}) with
$B=1$ is the well-known hedgehog configuration: \be
\label{hedgehog} U=e^{i \vec{\tau}\cdot \hat{r}F(r)}
 \ee
with $\tau_i =\lambda_i$ and $i \in \left\{1,2,3 \right\}$.  The
function $F(r)$ is a solution of the following radial equation
\cite{Adkins:1983ya}:
\begin{center}
\be
\begin{array}{c}
(\frac{1}{4} \tilde{r}^2 + 2 \sin^2{F}) F'' + \frac{1}{2} \tilde{r} F' + \sin{2F}F'^2
\\[.12in]
 - \frac{1}{4} \sin{2F} - \frac{\sin^{2}{F}
\sin{2F}}{\tilde{r}^2} = 0,
\end{array}
\ee
\end{center}
where we use the dimensionless variable $\tilde{r}=\frac{f_{\pi}}{\epsilon} r$.  The function $F(r)$ has the boundary conditions:
\[
F(r = 0) = \pi \;\;\; F(r \rightarrow \infty) \rightarrow 0.
\]
These boundary conditions ensure that $B=1$.

In the Skyrme model the angular momentum and isospin are
individually conserved. However, due to the $\vec{\tau}\cdot
\hat{r}$ structure of Eq. (\ref{hedgehog}), they are correlated
within the hedgehog ansatz.  This correlation leads to the
breaking of both rotational and isorotational invariance at the
level of the classical solution.  As discussed in Sec.~\ref{gen},
if one quantizes the collective degrees of freedom, the resulting
states will restore the broken symmetries and allow one to
identify well-defined physical states.

\section{Zero Modes}
\label{modes}

The goal of this section is to identify all of the static and
dynamical zero modes associated with flavor symmetry.  One way to
do this is to find explicit expressions for $\mathbb{A}$ and
$\mathbb{B}$ from Sec. ~\ref{gen} by expanding the full equation
of motion around a static hedgehog. One can then determine the
zero eigenmodes of $\mathbb{A}$ to obtain the static zero modes,
and then construct dynamical zero modes using Eq.~(\ref{dyn}).
However, this is not necessary.  One can use the known symmetry
properties of a given classical soliton to establish the static
zero modes. They are simply the infinitesimal changes in the
energetically flat directions which are allowed by the symmetries.

The dynamical zero modes are a bit more involved.  One can
consider a configuration which is slowly rotating  in the
direction of one of the static zero modes.  If the equation of
motion is satisfied to leading order in the rotation frequency
$\omega$, then in the language of Eq.~ (\ref{dyn}) we have
$\mathbb{B} \psi =0$ and we automatically have a dynamical zero
mode.  Conversely, if $\mathbb{B} \psi \ne 0$, then the equation
of motion will be violated by an amount proportional to $\omega
\mathbb{B} \psi$. To construct a dynamical zero mode in such a
case we need to invert the relation $\mathbb{B} \psi = \mathbb{A}
\Phi$ to find $\Phi$---provided it is invertible. In general this
requires knowledge of the detailed form of $\mathbb{A}$. However,
we know the preceding relation is not invertible and no dynamic
zero mode exists unless $\mathbb{B} \psi$ is orthogonal to
\emph{all} of the zero modes of $\mathbb{A}$.  As we will see, in
the Skyrme model all of the rotations either yield $\mathbb{B}
\psi =0$ and directly give a dynamic zero mode independent of the
form of $\mathbb{A}$, or they are not orthogonal to all of the
zero modes of $\mathbb{A}$, and thus one does not obtain a dynamical
zero mode. All dynamical zero modes in the Skyrme model can
therefore be found without explicitly constructing $\mathbb{A}$.

We begin our analysis of zero modes in the Skyrme model by embedding the $SU(2)$ hedgehog ansatz in
the up-down (u-d) subspace of $SU(3)$ \cite{Witten:1983tx}:
\be
 U_H=
\left(
\begin{array}{c|c}
 \exp{i (\vec{\tau} \cdot \hat{r})F(r)}& 0  \\
\hline 0 & 1  \\
\end{array}
\right).
\ee
 The hedgehog ansatz $U_H$ solves the static Skyrme
equation of motion (Eq. (\ref{eom})).  Note that the choice of the
u-d subspace embedding above was arbitrary since we are free to
choose any $SU(2)$ subspace.  As a result of this freedom, it can
be seen that a statically rotated hedgehog is also a solution to
the equation of motion, Eq. (\ref{eom}).  That is, the rotated
hedgehog below also solves the equation of motion:
 \be
 U = A U_H A^\dagger, A \in SU(3).
\ee
However, since $\lambda_8$ commutes
with $U_H$ we take
\[
A \in SU(3)/ U(1)_{\lambda_8}
\]
so that $A$ only depends on seven independent parameters.  This implies
there are seven static zero modes associated with such changes in
initial configuration.  They are given explicitly by
\be
\label{zm}
\delta U_{\rm s}(\lambda_j) = [-i\lambda_j,U_H] = \sin (F(r))[\lambda_j, \hat{r} \cdot \vec{\tau}]
\ee
for $j=1,\cdots, 7$.

We now analyze the dynamical zero modes.   We can construct rigidly
rotating hedgehogs by replacing $U_H$ with $A(t) U_H
A^{\dagger}(t)$, where $A(t) = \exp{i (\vec{\lambda} \cdot
\vec{\omega})t}$.  Here $(\vec{\lambda} \cdot \vec{\omega})$ is an
arbitrary linear combination of $SU(3)/ U(1)_{\lambda_8}$
generators, and $\omega$ is of order ${N_c}^{-1}$ and thus the
rotations are slow.  In the language of Sec. \ref{gen}, this form
automatically gives the dynamic zero modes if $\mathbb{B}=0$ and
gives us $\mathbb{B} \psi$ otherwise.  Without the WWZ term (that
is, for just Eq. (\ref{SK}) and $\mathbb{B}=0$), it can be seen by
direct substitution that $A(t) U_H A^{\dagger}(t)$ is an
$\mathcal{O}(\omega^2)$ approximate solution to the equation of
motion.

Now consider the effect of the WWZ term on dynamical rotations.  For
simplicity let $A(t)=\exp{(i\lambda_{p} \omega t)}$, where
$\lambda_p$ is one of the generators of $SU(3)/ U(1)_{{\lambda}_8}$.
When we substitute $U=A(t) U_H A^{\dagger}(t)$ into
$\Gamma_{wwz}=-i\epsilon^{\alpha \beta \gamma \nu} L_{\alpha}
L_{\beta} L_{\gamma} L_{\nu}$, we see that to first order in
$\omega$ \be
 \Gamma_{wwz}(\lambda_{p}) = \omega \epsilon^{i j k}\{[[\lambda_p,U_H]U_H^{\dagger},L_{i}], L_{j} L_{k}\}.
 \ee
In the preceding $\{\;,\;\}$ denotes an anticommutator.  Explicitly, we find:
\begin{widetext}
 \be
\begin{array}{ccc}
\label{gamma_res}
& \Gamma_{wwz}(\lambda_{1,2,3}) = 0 & \\
\Gamma_{wwz}(\lambda_4) = \omega ( a \lambda_4 - b \lambda_5 + c \lambda_6 - d \lambda_7) & & \Gamma_{wwz}(\lambda_5) = \omega ( b \lambda_4 + a \lambda_5 + d \lambda_6 + c \lambda_7)  \\
\Gamma_{wwz}(\lambda_6) =\omega (  c \lambda_4 + d \lambda_5 - a \lambda_6 - b \lambda_7) & & \Gamma_{wwz}(\lambda_7) =  \omega ( -d \lambda_4 + c \lambda_5 + b \lambda_6 + -a \lambda_7) , \\
\end{array}
\ee
where
\be
\begin{array}{cc}
\label{coeffs}
a  =  \frac{6 i}{r^2} \cos{\theta} \sin^{3}{\left(F(r)\right)} F'(r) &
b  =  \frac{48 i}{r^2} \cos^{2}{\left(\frac{F(r)}{2}\right)} \sin^{4}{\left(\frac{F(r)}{2}\right)} F'(r) \\
c  =  \frac{6 i}{r^2} \cos{\phi} \sin{\theta} \sin^{3}{\left(F(r)\right)} F'(r) &
d  =  \frac{6 i}{r^2} \sin{\phi} \sin{\theta} \sin^{3}{\left(F(r)\right)} F'(r). \\
\end{array}
\ee
\end{widetext}

>From Eqs.~(\ref{gamma_res}) we explicitly see that dynamically
rotating solutions generated by $\lambda_{4,5,6,7}$ do \emph{not}
satisfy the equations of motion, as the coefficients $a,b,c,d$ are
generally non-zero. On the other hand, dynamically rotating
solutions generated by $\lambda_{1,2,3}$ do satisfy the equations of
motion. In the approach of Sec. \ref{gen}, the right-hand sides of
Eqs.~(\ref{gamma_res}) are identified as $\mathbb{B} \psi$. The
construction in Eq.~(\ref{dyn}) can be used to produce dynamical
zero modes provided $\mathbb{B} \psi$ ({\it i.e.}
$\Gamma_{wwz}(\lambda_j)$) has no components in the directions of
the static zero modes.  For $j=1,2,3$ this is trivially true since
$\Gamma_{wwz}(\lambda_{1,2,3})=0$ and these three directions have
dynamical zero modes associated with them.  This is expected; the
WWZ term does not contribute to the equation of motion for these
modes since we effectively work in $SU(2)$ (see Ref.
\cite{Adkins:1983ya}).

On the other hand it is also straightforward
to see, by explicit computation, that all of the $\Gamma_{wwz}(\lambda_{4,5,6,7})$
have non-vanishing overlaps with the static zero modes of
Eq.~(\ref{zm}).  We define the overlap of two functions $f,g$ as
\be
\left\langle f|g\right\rangle = \frac{1}{2} \int{d^{3}x \; \mathrm{Tr}[f(x) g(x)]}.
\ee
With this definition, it can be shown that
\be
\begin{array}{c}
\left\langle \delta U_{\rm s}(\lambda_5)| \Gamma_{wwz}(\lambda_4) \right\rangle = \left\langle \delta U_{\rm s}(\lambda_7)| \Gamma_{wwz}(\lambda_6) \right\rangle = 9\pi^2
\\[.12in]
\left\langle \delta U_{\rm s}(\lambda_4)| \Gamma_{wwz}(\lambda_5) \right\rangle = \left\langle \delta U_{\rm s}(\lambda_6)| \Gamma_{wwz}(\lambda_7) \right\rangle = -9\pi^2 ,
\end{array}
\ee
and all other overlaps are zero.  This shows that each of the $\Gamma_{wwz}(\lambda_{4,5,6,7})$ has components along one of the static zero-mode directions.  Therefore we see that there are no collective dynamical zero modes along the $\lambda_{4,5,6,7}$ directions, {\it i.e.}, for dynamical rotations out of the u-d subspace.

The preceding demonstration shows that including the WWZ term in the
$SU(3)$ Skyrme Lagrangian eliminates four of the dynamical zero
modes. There are only three collective dynamical zero modes.  This is
significant since the standard treatment of three flavor hedgehog
models has been in the context of a rigid rotor treatment
\cite{Diakonov:1997mm, hedgehogs,Guadagnini}, which is based
implicitly on the assumption that there are seven collective dynamical zero
modes.

\section{Quantization}

There are ten total zero-modes (seven static and three dynamical)
that we can use as collective variables to quantize the soliton.
We will adopt the strategy of Sec. \ref{gen}.  We construct an
ansatz which allows for the most general collective motion
consistent with the zero mode structure.  This ansatz is labeled
by ten parameters (one for each zero mode) which will be promoted
to quantum operators. The appropriate ansatz for collective
rotations of the soliton is given by 
\be
 \label{ansatz} 
 A e^{i (\vec{\tau}
\cdot \vec{\omega} )t} U_H e^{-i (\vec{\tau} \cdot \vec{\omega})t}
A^{\dagger} 
\ee 
where $A \in SU(3) / U(1)_{\lambda_8}$, and
$(\vec{\tau} \cdot \vec{\omega})$ represents a linear combination
of the first three Gell-Mann matrices.  The seven parameters that
specify $A$, along with the three parameters that specify
$\vec{\omega}$, can now be used to collectively quantize the
Skyrmion following the procedure of Sec. \ref{gen}.

Our initial goal here is modest:  we seek only to characterize the
quantum numbers of the states that would result from such a
quantization procedure. In what follows we will obtain such a
characterization by using general group-theoretic arguments.   We
would like to identify the representations of $SU(3)$ that can arise as
a result of quantization. That is, we want to determine the $(p, q)$
that characterize these representations. To find these $(p,q)$ we
will focus on the maximum hypercharge in each representation.

The dimension of an $SU(3)$ representation is given by $2 Y_{max} +
1$, where $Y_{max}$ is the maximum hypercharge of the
representation. If we define $B$ as the baryon number and $S$ as the
strangeness, then the hypercharge for arbitrary $N_c$ is given by
$Y= \frac{N_c B}{3} + S$.  Note that the dimension of each
physically allowable representation scales as $N_{c}^1$ and becomes
infinite-dimensional in the large $N_c$ limit.  To make the problem
tractable we consider $N_c$ to be finite but arbitrarily large.

We now seek to determine $Y_{max}$ for the collectively rotated
hedgehog ($B=1$).  In Appendix \ref{noether}, we present an explicit
computation of $Y_{max}$ by constructing the Noether current
associated with the hypercharge. We find that at large $N_c$, \be
\label{ymax} Y_{max} = \frac{N_c}{3} \ee independently of the
parameters specifying the collective motion. By eliminating the
possibility of producing collective states with $Y>N_c/3$,  we see
that all physically allowable states must have $S \leq 0$.

Having found the dimensions of the physically allowable
representations, we can determine the $(p,q)$ that characterize
them.  We use Young tableaux to enumerate $SU(3)$ representations.
First, note that for $SU(3)$ we need only consider Young tableaux
with one or two rows.   Representations of the form $(p,q)$ are
known to have a maximum hypercharge given by $Y_{max} = 2q/3 + p/3$.
Equating this with the value in Eq.~(\ref{ymax}) yields $N_c= 2q +
p$ and implies that the number of boxes for any allowable representation
must be equal to $N_c$. Thus the relevant Young tableaux are those
that have the form $(p, \frac{N_c - p}{2})$:
\[
\yng(2,2)\cdots \yng(2,1)\;, \;\;\;\; \yng(2,2)\cdots \yng(3,1), \;\; \ldots \\\\.
\]

The hypercharge is maximized when $S=0$.  Let us focus on the
states with maximal hypercharge.  These
states correspond to the situation where each box in a tableaux
can be thought of as carrying either a $u$ or $d$ --- that is,
isospin $I=\pm 1/2$.  However, each of the $q$ double rows is an
anti-symmetric combination which contributes no net isospin. Thus
for the states of maximal hypercharge we have $I = p/2$. In the u-d
subspace, the Skyrmion is known to have $I=J$, implying that $J =
p/2$.  Using this information we can identify (by giving $(p,q),
\;Y, I$ and $J$) all of the physically allowable states resulting
from collective quantization of the Skyrmion.

\begin{table}[h,floatfix]

\begin{center}
\begin{tabular}{|c|c|c|}
\hline
~~~~~J=I~~~~~ & ~~~~~p~~~~~ &  ~Collective State~ \\ \hline\hline
$\frac{1}{2}$ & 1 & p,n \ldots  \\ \hline
$\frac{3}{2}$ & 3 & $\Delta$'s \ldots  \\ \hline
$\frac{5}{2}$ & 5 & Large $N_c$ artifact  \\ \hline
\end{tabular}
\caption{Allowable Representations}
\label{tab:AllowableRepresentations}
\end{center}
\end{table}
\noindent(In this model, states with $J>\frac{3}{2}$ are assumed
to be artifacts of the  large $N_c$ world.)

Note that these states are identical to those predicted in the
model-independent results of Ref. \cite{dashen}; they are also the
same as those obtained from the naive quark model.  Also observe
that while we find the same non-exotic states as predicted by
rigid rotor quantization, we do not find any exotic states with
$S>0$ such as the $\theta^{+}$ pentaquark.  This is one of the
principal differences between this systematic treatment of the
quantization and the {\it ad hoc} rigid rotor approach of Refs.
\cite{Guadagnini,hedgehogs}.

Finally let us construct the collective Hamiltonian.  This
can be done using the approach of Sec. \ref{gen}.  We insert the
ansatz Eq.~\ref{ansatz} into the full Skyrme Hamiltonian to obtain a collective
Hamiltonian. The parameters of the ansatz are again promoted to
collective quantum mechanical operators.  The collective
Hamiltonian is simply given by: \be \label{colham} H = M_0 +
\frac{1}{2} {\cal I} {\omega}^2 \, \ee where the mass $M_0$ and
the moment of inertia ${\cal I}$ are the standard Skyrmion
expressions \cite{Adkins:1983ya}.  Implementation of the procedure
in Sec. \ref{gen} for the Noether current for angular momentum
relates $\omega$ to the angular momentum $\vec{J} = {\cal I}
\vec{\omega}$. We arrive at the following expression for the
collective Hamiltonian: \be
\begin{array}{ccc}
 H & = & M_0 + \frac{\hat{\vec{J}}^2}{2{\cal I}} + {\cal O}({N_c}^{-2})
 \\
\Rightarrow E &= &  M_0 + \frac{j(j+1)}{2{\cal I}} + {\cal O}({N_c}^{-2}),
\label{ham}
\end{array}
\ee
where $\hat{\vec{J}}$ is the quantum angular momentum operator and $j$ is the spin of
the state.  This Hamiltonian gives the energy splittings for the
states given in Table \ref{tab:AllowableRepresentations}.

It is instructive to compare the collective Hamiltonian with that
obtained using the rigid rotor approach. A convenient way to write
the rigid rotor collective Hamiltonian is given in
Ref.~\cite{Diakonov:1997mm}: \be
\begin{array}{rll}
H & = & {M}_0 + \frac{1}{2 \cal{I}} \sum^{3}_{A=1} {\hat{J}_A}^2 + \frac{1}{2 I'} \sum^{7}_{A=1} {\hat{J}_A}^2
\\[.12in]
  & \Rightarrow & {M}_0 + \frac{1}{6 \cal{I}'} \left[ p^2 + q^2 + pq + 3(p+q) \right]
  \\[.12in]
  &+& \left(\frac{1}{2 \cal{I}}-\frac{1}{2 \cal{I}'}\right)j(j+1) - \frac{\left( N_c\right)^2}{24 \cal{I}'} + {\cal O}({N_c}^{-2})
  \, ,
\end{array}
\label{rr}
\ee
where $\hat{J}_A$ represents the generator of rotations in the
body-fixed frame. The system is characterized by two moments of
inertia: $\cal{I}$ is the moment of inertia in the u-d subspace
while ${\cal I}'$ characterizes rotations out of the u-d subspace.

Equation (\ref{rr}) disagrees in a critical way with
Eq.~(\ref{colham}): there are of two moments of inertia in
Eq.~(\ref{rr}) with one representing motion out of the u-d
subspace. This second moment of inertia is in fact not present in
the collective Hamiltonian (\ref{ham}), since there are no
dynamical zero modes for rotations out of the u-d subspace, and thus no
moment of inertia can be associated with such rotations.  Despite
this profound difference, the rigid rotor approach does capture
much of the correct physics.  In particular, if one uses the rigid
rotor expression in Eq.~(\ref{rr}) but restricts one's attention
to the physically allowable representations of Table
\ref{tab:AllowableRepresentations} with $q = \frac{N_c}{2} - J$ and
$p=2J$, the spectrum of the rigid rotor is found to be \be
 H = M_0 + \frac{j(j+1)}{2{\cal I}} + \frac{N_c}{4
\cal I}' + {\cal O}({N_c}^{-2}).
\ee
Note that this differs from
the correct spectrum only by the $\frac{N_c}{4 \cal I}'$ term,
which contributes an  overall constant at a subleading order
(relative order $1/N_c$).

\section{Conclusion}
We have shown how to collectively quantize the three-flavor
Skyrmion.  The basic approach can be used with minor modifications
for any hedgehog soliton model with three degenerate flavors. The
key to the approach is the distinction between dynamic and static
zero modes.  We have demonstrated explicitly that the
Witten-Wess-Zumino term eliminates four of the seven possible
dynamical zero modes. Using group-theoretic arguments together
with the zero mode results, we showed that the resulting states must
have $S \leq 0$, and must lie in representations characterized by
$(p,q)=(p, \frac{N_c - p}{2})$. To complete the specification of
the physically allowed collective states we demonstrated that
$J=p/2$. The $S \leq 0$ constraint implies that exotic baryons,
such as the $\theta^{+}$ ($1540$) pentaquark resonance, are not
predicted as collective states in the context of the $SU(3)$
flavor Skyrme model.

\begin{acknowledgements}
 The support of the U. S. Department of Energy through grant
number DE-FG02-93ER-40762 is gratefully acknowledged.  A.C. and T.R.D. gratefully acknowledge the support of the University of Maryland through the Senior Summer Scholar program.  We also thank Shmuel Nussinov for helpful discussions.
\end{acknowledgements}

\appendix
\section{The maximum hypercharge \label{noether}} In this appendix,
we will derive a formula for the hypercharge for collectively rotating hedgehogs in the Skyrme model, and will show that in the large $N_c$
limit it is bounded from above by $\frac{N_c B}{3}$.

To begin, observe that the action in Eq. (\ref{action}) is
invariant under rotations generated by $\lambda_8$.  We can
construct a conserved current associated with this symmetry.
Combining the expression for the current associated with the WWZ
term found by Witten \cite{Witten:1983tw} with the Noether current
obtained from the Skyrme Lagrangian (Eq. (\ref{SK})), we find \be
\label{current}
\begin{array}{l}
y^{\mu} = \frac{i}{2\sqrt{3}} \mathrm{Tr} [  -f^2_\pi \lambda_8 \left(L^\mu + R^\mu \right)
\\
- 2  \lambda_8 \epsilon^2 \left( [L_\nu,[L^\mu,L^\nu]] + [R_\nu,[R^\mu,R^\nu]]
\right)\\
 + \frac{N_c}{48\pi^2} \lambda_8  \epsilon^{\mu \nu \rho \sigma} \left( L_\nu L_\rho L_\sigma - R_\nu R_\rho R_\sigma \right) ],
\end{array}
\ee
where $R_\mu = U (\partial_\mu U^\dagger)$ and $L_\mu =
U^\dagger (\partial_\mu U)$.  The current $y^{\mu}$ is conserved
($\partial_{\mu} y^{\mu}=0$),  and the integral of $y^{0}$ is the
appropriately normalized hypercharge $Y$.

We now insert the ansatz for collective rotations (Eq.
(\ref{ansatz})) into Eq. (\ref{current}). Expanding to first order
in $\omega \sim 1/N_c$, we can write the hypercharge density as
$y_{0} = y_0^{(0)} + \omega y_0^{(1)}$, where
\begin{widetext}
\begin{eqnarray} y_0^{(1)}& = &  \frac{i}{2\sqrt{3}} N_c \mathrm{Tr} \left[
A^{\dagger} \lambda_8 A \left(  -\frac{f^2_\pi}{N_c}
[U^{\dagger}_H,\Omega] - \frac{2 \epsilon^2}{N_c} (
[\tilde{L}_k,[U^{\dagger}_H \Omega,\tilde{L}^k]]  - [
\tilde{R}_k,[\Omega U^{\dagger}_H,\tilde{R}^k] ) +
\frac{\epsilon^{i j k}}{48\pi^2} ([i(\vec{\tau} \cdot \hat{\omega}),
\tilde{L}_i \tilde{L}_j \tilde{L}_k  - \tilde{R}_i \tilde{R}_j
\tilde{R}_k] ) \right) \right]  \nonumber \\
  y_0^{(0)} & = &
\frac{i}{2\sqrt{3}} N_c \mathrm{Tr} [A^{\dagger} \lambda_8 A
\frac{\epsilon^{i j k} }{48\pi^2}  ( \tilde{L}_i \tilde{L}_j
\tilde{L}_k  - \tilde{R}_i \tilde{R}_j \tilde{R}_k) ],
\end{eqnarray}
\end{widetext}
with $\tilde{R}_i = U_H (\partial_i U_H^\dagger)$, $\tilde{L}_i =
U_H^\dagger (\partial_i U_H)$ and $\Omega=[i(\vec{\tau} \cdot
\hat{\omega}), U_H]$.

The hypercharge of the collectively rotating hedgehog Skyrmion,
$Y_{\mathrm{col}}$, is given by
 \be \label{hypercharge}
Y_{\mathrm{col}} = \int d^{3}x\; y_0. \ee If we consider only static
rotations within the u-d subspace ($\omega = 0$ and $A$ in the
u-d subspace of $SU(3)$), we find that the hypercharge is given by
 \be
Y_{\mathrm{col}} = \frac{1}{2\sqrt{3}} N_c \left(\frac{i
\epsilon^{i j k}}{24\pi^2} \int {d^{3}x\;  \mathrm{Tr} [ \lambda_8
( \tilde{L}_i \tilde{L}_j \tilde{L}_k)]}\right). 
\ee 
Since the expression in parentheses is proportional to the baryon number $B$
defined in Eq. (\ref{baryonNum}) (specifically, it is just
$\frac{2 B}{\sqrt{3}}$), we obtain the relation $Y_{\mathrm{col}}
= \frac{ N_c B}{3}$ for the hypercharge of a hedgehog Skyrmion undergoing
static rotations in the u-d subspace.

We now claim that in the large $N_c$ limit
$Y_{\mathrm{col}}=\frac{ N_c B}{3}$ is an upper bound. To see
this, let us relax the preceding restrictions and consider a
hedgehog that is undergoing slow ($\omega \sim 1/N_c$) collective
dynamical rotations, and allow general $SU(3)$ static rotations.
After the spatial integral in Eq. (\ref{hypercharge}) is
evaluated, only a term proportional to $\hat{\omega}\cdot
\vec{\tau}$ can survive in the term coming from $y_0^{(1)}$. To
see this, observe that integrating $y_0^{0}$ over space is
equivalent to integrating over isospace due to the correlation of
space and isospace in a hedgehog embedded in the u-d space;  the
only direction that can remain is the one set by $\vec{\tau} \cdot
\hat{\omega}$. Integrating $y_0^{(0)}$ over space yields
\begin{equation*}
 \frac{N_c B}{3} \mathrm{Tr} \left[A^{\dagger} \lambda_8 A
\left(
\begin{array}{ccc}
1 & 0 & 0 \\
0 & 1 & 0 \\
0 & 0 & 0 \\
\end{array}
\right)\right] = \frac{N_c B}{3} \mathrm{Tr}[A^{\dagger} \lambda_8 A (\frac{2}{3}\mathbf{1} +\lambda_8/\sqrt{3})].
\end{equation*}
Thus we see that the hypercharge of a
collectively rotating hedgehog ($B=1$) is given by
\be
Y_{\mathrm{col}}  = a N_c \omega \mathrm{Tr} \left[A^\dagger
\lambda_8 A (\hat{\omega}\cdot \vec{\tau}) \right] + \frac{N_c }{3}
\mathrm{Tr} \left[
    A^{\dagger} \lambda_8 A
(\frac{2}{3}\mathbf{1} +\lambda_8/\sqrt{3})\right] ,
\ee
 where $a$ is a real coefficient of order $N_c^0$ that is
proportional to the moment of inertia $\cal{I}$ of an $SU(2)$ hedgehog: $a N_c = \frac{4}{\sqrt{3}}
{\cal I}$.

Since $Y_{\mathrm{col}}$ is an isoscalar, the orientation of
$\vec{\omega}$ is irrelevant, and thus without loss of generality we
set $\vec{\omega}= \omega \hat{z}$.  We can write $A^\dagger
\lambda_8 A = \sum^{8}_{i=1} c_i \lambda_i$ with the condition that
$\sum^{8}_{i=1} \left| c_i \right|^2 = 1$.  Since a unit matrix in
the u-d subspace of $SU(3)$ can be written as
$\frac{2}{3}\mathbf{1}+ \frac{1}{\sqrt{3}}\lambda_8$, we can
rewrite the hypercharge as
\be
 Y_{\mathrm{col}} = 2 a \omega N_c c_3
+ \frac{N_c}{3} \sqrt {1 - \sum^{7}_{i=1} \left| c_i \right|^2 }.
\ee 
Maximizing the hypercharge with respect to the condition
$\sum^{8}_{i=1}{ \left| c_i \right|^2} = 1$, we see that $ c_3 = 6
a \omega \sim 1/N_c$ and $c_i = 0$ for all $i \neq 3,8$.  Thus in
the large $N_c$ limit $c_3$ vanishes.  This leads to the condition
$\left| c_8 \right|^2 = 1$ and $Y_{\mathrm{col}} = \frac{N_c}{3}$.
This shows that in the large $N_c$ limit, $Y_{\mathrm{col}} =
\frac{N_c}{3}$ is a global maximum for the hypercharge of a
collectively rotating hedgehog Skyrmion.\\

\bibliographystyle{amsplain}

\end{document}